# A Firefly Algorithm-based Spectral Fitting Technique for Wavelength Modulation Spectroscopy Systems

Tingting Zhang, Yongjie Sun, Pengpeng Wang, Yufeng Qiu, Chenxi Wang, Xiaohui Du, Shaokai Li, Haixu Liu, Tongwei Chu, Cunguang Zhu

***Abstract*—This paper proposes a novel calibration-free wavelength modulated spectroscopy (WMS) spectral fitting technique based on the firefly algorithm. The technique by simulating the information interaction behavior between fireflies to achieve the retrieval of gas concentration and laser parameters. Contrasted with the spectral fitting technique based on the classical Levenberg-Marquardt (LM) algorithm, the retrieval of gas concentrations by this technique is weakly dependent on the pre-characterization of the laser parameters. We select the P(13) absorption line of $C_2H_2$ at 1532.82 nm as the target spectra and compare the performance of two optimization method (LM and firefly) on gas concentration and laser parameters retrieval by simulation. The simulation results show that the spectral fitting technique based on the firefly algorithm performs better in terms of convergence speed and fitting accuracy, especially in the multi-parameter model without exact characterization.**

## I. INTRODUCTION

With the rapid development of infrared diode laser technology, tunable diode laser absorption spectroscopy (TDLAS) has been widely used in greenhouse gas and air pollution monitoring, combustion monitoring, industrial process control, and other fields [1]-[6]. Wavelength modulation spectroscopy (WMS), a branch of TDLAS, has been widely used for in-situ measurements in harsh environments such as strong turbulence, high temperature or high pressure due to its high sensitivity and robustness to background noise [7]-[16].

While WMS technique offers many benefits, quantitative WMS measurements are still challenging due to the complexity of the WMS signal. The WMS measurement is performed using the harmonics of the gas absorption spectrum, which depends on the laser intensity and the line-shape. Laser intensity is susceptible to interference from beam steering, scattered particles and window fouling. Moreover, the line-shape strongly dependents on temperature (Doppler broadening) and pressure (Collision broadening). The traditional approach chosen by the researchers was to perform in-situ calibration with a known gas mixture [17]. However, this is impractical when gas conditions of interest are poorly understood or highly transient.

Consequently, several researchers have developed calibration-free WMS methods [18]-[22]. Li et al. [18] proposed an analytical model that simulates WMS signals as a function of the well-characterized laser parameters and absorption spectra. Rieker et al. [19] utilized the second harmonic (2*f*) normalized by the first harmonic (1*f*) to infer the gas parameters, thereby eliminating the effect of light intensity fluctuations. Sun et al. [20] developed a calibration-free method using the measured non-absorption laser intensity without the need to build an analytical model to describe the temporal laser intensity variations. Based on Sun's work, Christopher et al. [21] proposed a strategy that enables accurate calibration-free WMS measurements of gas properties without needing prior knowledge of the transition line-shape parameters. Yang et al. [22] proposed a gas concentration retrieval approach based on the first harmonic phase angle ($\theta_{1f}$) method, which is immune to the laser intensity and the demodulation phase. A second harmonic phase angle method ($\theta_{2f}$) based on WMS was subsequently proposed for trace gas detection enabling background-free detection and immunity to light intensity fluctuations [23]. In the past decades of calibration-free WMS method development, Levenberg-Marquardt(LM) algorithm has been widely used to deal with nonlinear least-squares problems in spectral fitting for gas concentration retrieval [21], [24]-[30]. For example,

This work was supported by the Natural Science Foundation of China (Grant No. 61705080), the Promotive Research Fund for Excellent Young and Middle-Aged Scientists of Shandong Province (Grant No. ZR2016FB17), the Scientific Research Foundation of Liaocheng University (Grant No. 318012101), and the Startup Foundation for Advanced Talents of Liaocheng University (Grant No. 318052156 and 318052157). (Tingting Zhang and Yongjie Sun contributed equally to this work.) (Corresponding author: Cunguang Zhu.)
Tingting Zhang, Yufeng Qiu, Pengpeng Wang, Chenxi Wang, Xiaohui Du, Shaokai Li, Haixu Liu, and Cunguang Zhu are with the school of Physics Science and Information Technology, Liaocheng University, Liaocheng, 252000, China (e-mail: zhtingting202006@163.com; wangpengpeng@lcu.edu.cn; qyf793372044@163.com; w15216453338@163.com; DD15169505459@163.com; 13176792668@163.com; xs245110842@163.com; cunguang_zhu@163.com).
Yongjie Sun is with the school of Physics and Technology, University of Jinan, Jinan, 250024, China (e-mail: 202021200721@stu.ujn.edu.cn).
Tongwei Chu is with the Department of Electrical and Computer Engineering, Duke University, Durham, NC 27710, USA (e-mail: sps_chutw@163.com).

Yang et al. [24] utilized the LM algorithm fitting the $2f$ spectrum of known concentration methane to the measured spectrum for measuring low concentration methane accurately. Cui et al. [30] developed a ppm-level CO sensor system for a sulfur hexafluoride decomposition analysis in power systems by fitting the $2f$ spectral information using the LM algorithm. Li et al. [25] developed a single continuous wave room-temperature quantum cascade laser sensor that used the LM algorithm for data processing to achieve simultaneous concentration measurements of atmospheric carbon monoxide, nitrous oxide, and water vapor. However, the LM algorithm required the support of an accurate pre-characterization due to its needed starting values of the free parameters. When multiple free parameters are included in the model, a matrix of high-order coefficients is generated, leading to high computational dimensionality, so its optimization results often do not match expectations.

In this paper, a calibration-free WMS spectral fitting technique based on the firefly algorithm (FA) is proposed, which simulates the information interaction behavior between fireflies to retrieve gas concentration and laser parameters. Contrasted with the spectral fitting technique based on the classical LM algorithm, the retrieval of gas concentrations by this technique weakens the dependence of gas concentration retrieval on pre-characterization. The simulation results show that the spectral fitting technique based on the firefly algorithm performs better in terms of convergence speed and fitting accuracy, especially in the multi-parameter model without exact characterization.

## II. THEORY AND METHODOLOGY

### A. Theory

In a scanned WMS-$2f/1f$ system, the laser is typically driven by a low-frequency sinusoidal scanning current superimposed on a high-frequency sinusoidal modulation current. Both the frequency and intensity of the outgoing laser will change, with a phase delay between them, which can be expressed as

$$I_0(t) = \overline{I_0}(v_c)[1 + i_0\cos(\omega t + \varphi_1) + i_2\cos(2\omega t + \varphi_2)] \quad (1)$$

$$v(t) = v_c + \Delta v \cdot \cos(\omega t) \quad (2)$$

where $\overline{I_0}(v_c)$ is the average laser intensity at the center laser frequency $v_c$, $i_0$ and $i_2$ are the linear and nonlinear intensity modulation (IM) depth (normalized by the average laser intensity), modulation angular frequency $\omega = 2\pi f_m t$, $f_m$ is laser modulation frequency (FM), $\varphi_1$ and $\varphi_2$ are the phase shift between FM and linear and nonlinear IM, respectively, $\Delta v$ is the modulation depth of the laser frequency.

According to Beer-Lambert law, when the laser beam passes through a gas cell filled with the absorbing gas, the transmitted laser intensity at frequency $v$ can be expressed as

$$I_t(t) = I_0(t) \cdot \exp[-\alpha(v(t))] \quad (3)$$

where $I_0(t)$ is incident laser intensity, $\alpha(v(t))$ is the spectral absorbance, in the case of weak absorption ($\alpha < 0.05$),

$$I_t(t) = I_0(t) \cdot \exp[-\alpha(v(t))] \approx I_0(t) \cdot [1 - \alpha(v(t))]$$
$$= I_0(t) \cdot [1 - PS(T)CLg(v, v_0)] \quad (4)$$

where $P$ is the total pressure of the mixed gas species, $S(T)$ is the line strength of the absorption spectrum at temperature $T$, $C$ is the concentration of the gas to be measured, $L$ is the length of the effective absorption path, $g(v,v_0)$ is the line shape function at optical frequency $v$ of the absorption feature, $v_0$ is the line-center frequency of the absorption spectrum.

At atmospheric pressure, $g(v,v_0)$ can be expressed as Lorentzian line shape function

$$g(v,v_0) = \frac{2}{\pi \Delta v_c} \frac{1}{1+[x+m\cos(\omega t)]^2} \quad (5)$$

$$x = 2\frac{v_c - v_0}{\Delta v_c} \quad (6)$$

$$m = \frac{2\Delta v}{\Delta v_c} \quad (7)$$

where $\Delta v_c$ is the full width at half-maximum, $x$ is the normalized frequency, $m$ is the modulation index.

when the laser is modulated by a sinusoidal injection current, $\exp[-\alpha(v(t))]$ can be expanded in a Fourier cosine series as follows

$$\exp[-\alpha(v(t))] = \sum_{k=0}^{\infty} H_k(v_c, \Delta v) \cdot \cos(k\omega t) \quad (8)$$

here $H_k$ is the kth Fourier coefficient and can be expressed as

$$H_0(v_c, \Delta v) = \frac{1}{2\pi}\int_{-\pi}^{\pi} \exp[-\alpha(v(t))]d\theta \quad (9)$$

$$H_k(v_c, \Delta v) = \frac{1}{\pi}\int_{-\pi}^{\pi} \exp[-\alpha(v(t))] \cdot \cos(k\theta)d\theta \quad (10)$$

The transmitted intensity $I_t(t)$ is multiplied by the quadrature reference signal $\cos(\omega t)$ and $\cos(\omega t + \pi/2)$ in the lock-in amplifier (LIA), and the X and Y components of the $1f$ signal are obtained after low-pass filtering as follows

$$X_{1f} = \frac{\overline{I_0}}{2}\left[H_1 + i_0(H_0 + \frac{H_2}{2})\cos\varphi_1 + \frac{i_2}{2}(H_1 + H_3)\cos\varphi_2\right] \quad (11)$$

$$Y_{1f} = \frac{\overline{I_0}}{2}\left[i_0(H_0 - \frac{H_2}{2})\sin\varphi_1 + \frac{i_2}{2}(H_1 - H_3)\sin\varphi_2\right] \quad (12)$$

the X and Y components of the $2f$ signal are given by

$$X_{2f} = \frac{\overline{I_0}}{2}\left[H_2 + \frac{i_0}{2}(H_1 + H_3)\cos\varphi_1 + i_2 H_0 \cos\varphi_2\right] \quad (13)$$

$$Y_{2f} = \frac{\overline{I_0}}{2}\left[\frac{i_0}{2}(H_1 - H_3)\sin\varphi_1 + i_2 H_0 \sin\varphi_2\right] \quad (14)$$

then the first harmonic normalized second harmonic $S_{2f/1f}$ are defined as

$$S_{2f/1f} = \sqrt{(\frac{X_{2f}}{X_{1f}})^2 + (\frac{Y_{2f}}{Y_{1f}})^2}$$
$$= \sqrt{\left(\frac{H_2 + \frac{i_0}{2}(H_1+H_3)\cos\varphi_1 + i_2 H_0\cos\varphi_2}{H_1 + i_0(H_0+\frac{H_2}{2})\cos\varphi_1 + \frac{i_2}{2}(H_1+H_3)\cos\varphi_2}\right)^2 + \left(\frac{\frac{i_0}{2}(H_1-H_3)\sin\varphi_1 + i_2 H_0\sin\varphi_2}{i_0(H_0-\frac{H_2}{2})\sin\varphi_1 + \frac{i_2}{2}(H_1-H_3)\sin\varphi_2}\right)^2} \quad (15)$$

Which states that $S_{2f/1f}$ signal is a function of multiple parameters, including gas concentration $C$, gas absorption line shape $g(v,v_0)$, Fourier expansion coefficient $H_k$, the linear and nonlinear IM depth $i_0$ and $i_2$, and the phases shift $\varphi_1$ and $\varphi_2$ between FM and linear and nonlinear IM, respectively.

Therefore, both gas concentrations and laser parameters are then inferred from the spectral fitting of the $S_{2f/1f}$ signal.

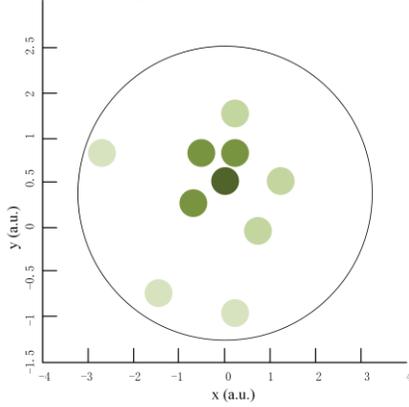

(a) Initial locations of firefly population

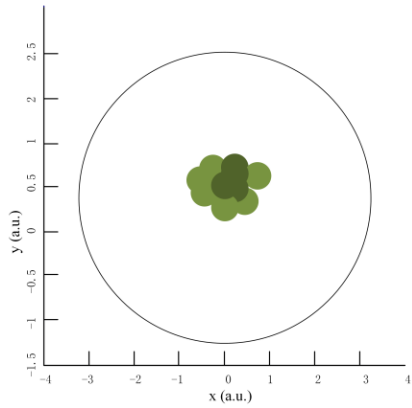

(b) Locations after optimization

Fig. 1. The FA optimization schematic.

### B. FA-based calibration-free WMS spectral fitting technique

In this paper, we developed an FA-based spectral fitting technique that simulates the information interaction behavior between fireflies to perform calibration-free measurements of gas concentration and laser parameters.

Firefly individuals look for their partners in the visual area according to their luminous behaviors and move toward brighter (better positioned) fireflies. The luminosity of firefly is related to the objective function. When the termination condition is satisfied, the highest luminosity value of the firefly is the optimal solution for the objective function. The FA optimization schematic is shown in Fig. 1. Fig. 1(a) is the initial location of the firefly population; Fig. 1(b) is the location of the optimized fireflies. As shown in the figure, each green circle indicates a firefly and the depth of green represents different luminosity levels. The center of the black circle in the figure is the highest global luminosity point also indicates the optimal solution of the objective function.

The FA-based calibration-free WMS spectral fitting technique flow chart is shown in Fig. 2, the specific processes are as follows. The free parameters of the fitting procedure may include the modulation index $m$, gas concentration $C$, the linear and nonlinear IM depth $i_0$ and $i_2$ and the phases shift $\varphi_1$ and $\varphi_2$ between FM and linear and nonlinear IM, respectively. We denote the free parameters by the vector $\beta$ that defines the current position of the firefly.

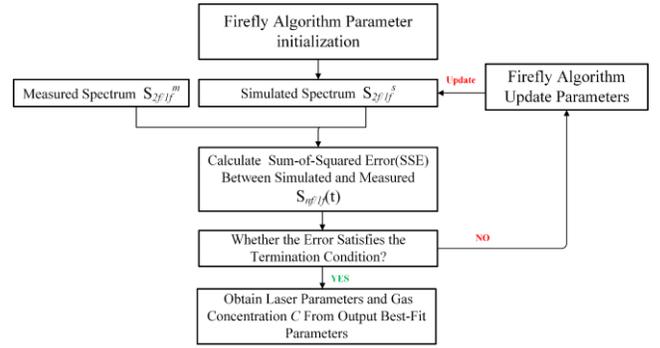

Fig. 2. Flow chart for seeking the optimal solution of gas properties and laser parameters using the FA.

Step 1: Parameter initialization - Generate $W$ vectors $\beta_i$ for the fitting procedure, which are randomly assigned within a specific range, $i$= 1, 2, … , $W$.

Step 2: Acquisition of simulated spectra - The simulated spectra $f_i(x_k)$ are obtained by substituting the initialized parameters $\beta_i$ into Eq. (15), $x_k$ is the normalized frequency, $k$= 1, 2, 3, …, $N$, $N$ is the total number of sampling points in the simulated spectra.

Step 3: Parameters updating - The objective function $F$ is used to evaluate the luminosity of each firefly, which is negatively correlated with each other. The objective function $F(\beta_i)$ is defined as follows:

$$\text{Given } F : R^D \to R$$
$$\text{Find}$$
$$z_k = y_k - f_i(x_k, \beta_i)$$
$$F(\beta_i) = \sum_{k=1}^{N} z_k^2$$
$$\text{End}$$

(16)

where $R^D$ represents the $D$-dimensional solution space, $D$ is dependent on the number of free parameters, $y_k$ is measured spectra, $z_k$ is the residual between the $k$th sampling points of measured spectra and simulated spectral, and the objective optimization function $F(\beta_i)$ is the sum of squares of all residuals.

Step 4: Judgment - If the convergence between measured spectra $S^m_{2f/1f}$ and simulated spectral $S^s_{2f/1f}$ does not satisfy the termination condition of the optimization, then continue to Step 5; Otherwise, perform Step 6.

Step 5: Firefly movement - Fireflies rely on position and luminosity updates to complete optimization, which determines the search route of fireflies in space. The firefly $i$ with less luminosity value is attracted to and approached by the brighter firefly $j$, the attractiveness function of the firefly is established by

$$h(r) = h_0 e^{\gamma r_{ij}^2}$$

(17)

where $r_{ij}$ is the distance from $x_i$ and $x_j$, $h_0$ is the firefly attractiveness value at $r = 0$ and $\gamma$ is the light absorption coefficient, generally set as a constant, $r_{ij}$ can be defined as

$$r_{ij} = \|x_i - x_j\| = \sqrt{\sum_{d=1}^{D}(x_{id} - x_{jd})^2}$$

(18)

TABLE 1
SUMMARY OF SPECTRAL PARAMETERS

| Symbol | Quantity | Value |
|---|---|---|
| $i_0$ | linear IM depth at line-center frequency | 0.15 |
| $i_2$ | nonlinear IM depth at line-center frequency | $3 \times 10^{-3}$ |
| $\varphi_1$ | phases shift between FM and linear IM | $0.6\pi$ rad |
| $\varphi_2$ | phases shift between FM and nonlinear IM | $0.5\pi$ rad |
| $m$ | modulation index | 1.5 |
|  | isotopologue | $^{12}C_2H_2$ |
| $v_c$ | line-center frequency of transition | 6523.8792 cm$^{-1}$ |
| $\Delta v_c/2$ | half width at half-maximum | 0.0777 cm$^{-1}$ |
| $S$ | line strength | $1.035 \times 10^{-20}$ cm/mol |
| T | arbitrary temperature | 296 K |
| P | total pressure | 1 atm |
| L | absorption path length | 20 cm |

where $d$ indicates the $d$th free parameters, $d= 1, 2, \ldots, D$, firefly $i$ will move towards firefly $j$, and the position of firefly $i$ is updated by the following equation

$$x_i^{t+1} = x_i^t + h_0 e^{\gamma r_{ij}^2}(x_i^t - x_j^t) + \lambda \left[ rand - \frac{1}{2} \right] \quad (19)$$

$\lambda \in [0,1]$ is the step factor, $rand \in [0,1]$ is a random value.

The brightest firefly moves randomly, which is defined as

$$x_i^{t+1} = x_i^t + \lambda rand() \quad (20)$$

After updates of individual firefly luminance and positions in steps 3 and 5, the firefly population will keep moving closer to the region of highest luminance.

Step 6: Termination condition - If the convergence between measured spectra $S^m_{2f/1f}$ and simulated spectra $S^s_{2f/1f}$ satisfies the termination condition of the optimization, the value of best-fit parameters is output at this point, and laser parameters and gas concentration $C$ are obtained.

### III. SIMULATION

The feasibility of the FA-based WMS spectral fitting technique is verified in the following. In order to avoid the characterization errors introduced by the metrology instruments affecting the evaluation of the algorithm performance, we chose to perform the validation by simulation in MATLAB R2019b platform. We compared the spectral fitting effect based on the LM algorithm with that based on the FA under the same conditions.

The P(13) absorption line of $C_2H_2$ at 1532.82 nm was selected as the target spectra for the following simulations. Given the values of the spectral parameters in Table 1, the measured spectra that should have been collected in the experiment were replaced by the simulated data set calculated by Eq. (15), which can be expressed as

$$[X_{4000}, Y_{4000}] = [(x_1, y_1), (x_2, y_2), \ldots, (x_{4000}, y_{4000})]$$

The above virtual measured spectrum is first fitted using the LM algorithm, which is discussed in two cases. The first case is setting only the gas concentration $C$ as a free parameter and all other parameters as known parameters, and the results are shown in Fig. 3. Fig. 4 shows the fitting results for the second case, i.e., more parameters are set as free parameters, including optical frequency modulation index $m$, gas concentration $C$, the linear and nonlinear IM depth $i_0$ and $i_2$, the phases shift $\varphi_1$ and $\varphi_2$ between FM and linear and nonlinear IM, respectively. In both cases, the convergence time is limited to 150 s. Comparing Fig. 3 and Fig. 4, we can observe that the residuals increase by 13 orders of magnitude ($10^{-16}$ to $10^{-3}$) when the number of free parameters increases from 1 to 6. The fitting effect of the LM algorithm deteriorates significantly with the increase in the number of free parameters.

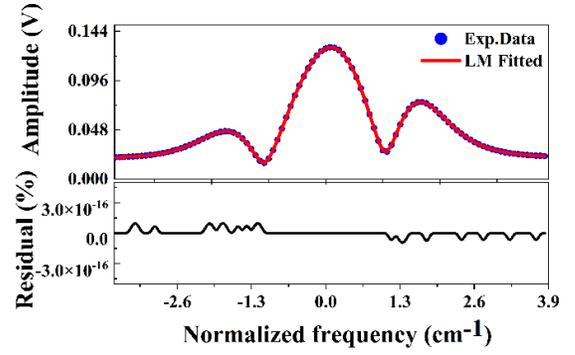

Fig. 3. Measured and fitted WMS-$2f/1f$ signals of free parameter $\beta_1=[C]$ with the LM algorithm convergence time of 150 s.

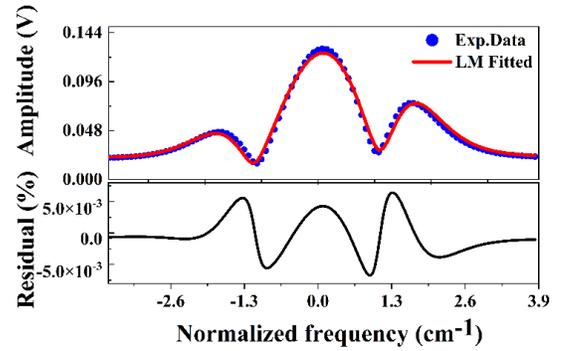

Fig. 4. Measured and fitted WMS-$2f/1f$ signals of free parameter $\beta_2=[m, C, i_0, i_2, \varphi_1, \varphi_2]$ with the LM algorithm convergence time of 150 s.

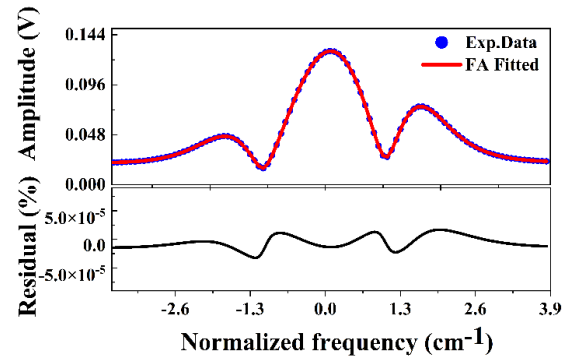

Fig. 5. Measured and fitted WMS-$2f/1f$ signals of free parameter $\beta_2=[m, C, i_0, i_2, \varphi_1, \varphi_2]$ with the FA convergence time of 30 s.

To contrast with the LM algorithm, the FA is used to fit the same virtual measured spectrum and free parameters as the LM algorithm, in which convergence time was reduced to 30 s. We keep the free parameters the same as in Fig. 4, and the fitting results are shown in Fig. 5, where residuals decrease by two orders of magnitude ($10^{-3}$ to $10^{-5}$), in which convergence time is only one-fifth of the LM algorithm.

TABLE 2
THE BEST-FIT FREE PARAMETERS PREDICTED BY THE LM ALGORITHM AFTER AN CONVERGENCE TIME OF 150 S

| Free parameters | Expected value | Predicted of the LM algorithm | Relative Errors |
|---|---|---|---|
| $m$[cm$^{-1}$] | 1.5000 | 1.5904 | 6.03% |
| $c$[ppmv] | 355.2 | 408.6 | 15.03% |
| $i_0$[pm/mA] | 0.1500 | 0.1386 | 7.60% |
| $i_2$[pm/mA] | 0.0030 | 0.0032 | 6.67% |
| $\varphi_1$[π rad] | 0.6000 | 0.6309 | 5.15% |
| $\varphi_2$[π rad] | 0.5000 | 0.4747 | 5.06% |

TABLE 3
THE BEST-FIT FREE PARAMETERS PREDICTED BY THE FA AFTER AN CONVERGENCE TIME OF 30 S

| Free parameters | Expected value | Predicted of the FA | Relative Errors |
|---|---|---|---|
| $m$[cm$^{-1}$] | 1.5000 | 1.5026 | 0.17% |
| $c$[ppmv] | 355.2 | 354.8 | 0.11% |
| $i_0$[pm/mA] | 0.1500 | 0.1526 | 1.73% |
| $i_2$[pm/mA] | 0.0030 | 0.0029 | 3.33% |
| $\varphi_1$[π rad] | 0.6000 | 0.6045 | 0.75% |
| $\varphi_2$[π rad] | 0.5000 | 0.5017 | 0.34% |

Table 2 and Table 3 show the best-fit free parameters of Fig. 4 and Fig. 5, respectively. It is evident that the relative errors of free parameters predicted by the LM algorithm all exceed 5%. However, the fitting effect of the FA is significantly improved compared to the LM algorithm. For instance, the linear and nonlinear IM depth $i_0$ and $i_2$ are consistent with predicted values within 5%; the rest of the parameters are consistent with predicted values within 1%.

In summary, the FA outperforms the LM algorithm in terms of convergence time and error for the model with multiple free parameters. Satisfactory multi-parameter optimization can get rid of the dependence on pre-characterization and also prevent the measurement error caused by pre-characterization failure.

## IV. CONCLUSION

This paper develops a novel calibration-free WMS spectral fitting technique based on the FA to retrieve gas concentration and laser parameters by simulating the information interaction behavior between fireflies. Contrasted with the spectral fitting technique based on the classical LM algorithm, the retrieval of gas concentrations by this technique is weakly dependent on the pre-characterization of the laser parameters. Our simulation results show that the relative errors of best-fit parameters predicted by the LM algorithm all exceed 5%, but most errors predicted by the FA are within 1%. All these results prove that the FA outperforms the LM algorithm in terms of convergence time and error for the model with multiple free parameters. This technique weakens the dependence of gas concentration retrieval on pre-characterization. The influence of pre-characterization failure caused by temperature change and laser aging on the measurement is effectively avoided.

## REFERENCES

[1] H. Nasim, Y. Jamil, "Recent advancements in spectroscopy using tunable diode lasers," *Laser Phys. Lett.*, vol. 10, no. 4, pp. 043001-043014, Feb. 2013.
[2] G. Galbács, "A review of applications and experimental improvements related to diode laser atomic spectroscopy," *Appl. Spectrosc. Rev.*, vol. 42, no. 3, pp. 259-303, Feb. 2007.
[3] K. Song, and E. C. Jung, "Recent developments in modulation spectroscopy for trace gas detection using tunable diode lasers," *Appl. Spectrosc. Rev.*, vol. 38, no. 4, pp. 395-432, Aug. 2006.
[4] S. Reuter, J. S. Sousa, G. D. Stancu, and J. H. V. Helden, "Review on VUV to MIR absorption spectroscopy of atmospheric pressure plasma jets," *Plasma Sources Sci. Technol.*, vol. 24, no. 5, pp. 054001-054041, Aug. 2015.
[5] T. Fernholz, H. Teichert, V. Ebert, "Digital, phase-sensitive detection for in situ diode-laser spectroscopy under rapidly changing transmission conditions," *Appl. Phys. B.*, vol. 75, no. 2, pp. 229-236, Sep. 2002.
[6] J. Liu, "Near-infrared diode laser absorption diagnostic for temperature and water vapor in a scramjet combustor," *Appl. Opt.*, vol. 44, no. 31, pp. 6701-6711, 2005.
[7] L. C. Philippe, and R. K. Hanson, "Laser diode wavelength-modulation spectroscopy for simultaneous measurement of temperature, pressure, and velocity in shock-heated oxygen flows," *Appl. Opt.*, vol. 32, no. , pp. 6090-6103, 1993.
[8] C. S. Goldenstein, R. M. Spearrin, I. A. Schultz, J. B. Jeffries, and R. K. Hanson, "Wavelength-modulation spectroscopy near 1.4 μm for measurements of $H_2O$ and temperature in high-pressure and-temperature gases," *Meas. Sci. Technol.*, vol. 25, no. 5, pp. 055101-055109, Mar. 2014.
[9] R. M. Spearrin, C. S. Goldenstein, J. B. Jeffries, and R. K. Hanson, "Quantum cascade laser absorption sensor for carbon monoxide in high-pressure gases using wavelength modulation spectroscopy," *Appl. Opt.*, vol. 53, no. 9, pp. 1938-1946, Mar. 2014.
[10] W. Cai, and C. F. Kaminski, "A tomographic technique for the simultaneous imaging of temperature, chemical species, and pressure in reactive flows using absorption spectroscopy with frequency-agile lasers," *Appl. Phys. Lett.*, vol. 104, no. 3, pp. 034101-1-034101-5, Jan. 2014.
[11] S. Neethu, R. Verma, S. S. Kamble, J. K. Radhakrishnan, P. P. Krishnapur, V. C. Padaki, "Validation of wavelength modulation spectroscopy techniques for oxygen concentration measurement," *Sens. Actuators B Chem.*, vol. 192, pp.70-76, 2014.
[12] Z. Peng, Y. Ding, L. Che, X. Li, and K. Zheng, "Calibration-free wavelength modulated TDLAS under high absorbance conditions," *Opt. Express.*, vol. 19, no. 23, pp. 23104-23110, Oct. 2011.
[13] T. Cai, G. Gao, and M. Wang, "Simultaneous detection of atmospheric $CH_4$ and CO using a single tunable multi-mode diode laser at 2.33 μm," *Opt. Express.*, vol. 24, no. 2, pp. 859-873, Jan. 2016.
[14] L. Shao, B. Fang, F. Zheng, X. Qiu, Q. He, J. Wei, C. Li, W. Zhao, "Simultaneous detection of atmospheric CO and $CH_4$ based on TDLAS using a single 2.3 μm DFB laser," *Spectrochim. Acta A Mol. Biomol. Spectrosc.*, vol. 222, pp. 117118-117123, Nov. 2019.
[15] Q. He, P. Dang, Z. Liu, C. Zheng, Y. Wang, "TDLAS–WMS based near-infrared methane sensor system using hollow-core photonic crystal fiber as gas-chamber," *Opt. Quant. Electron.*, vol. 49, no. 3, pp. 1-11, Feb. 2017.
[16] D. S. Bomse, A. C. Stanton, and J. A. Silver, "Frequency modulation and wavelength modulation spectroscopies: comparison of experimental methods using a lead-salt diode laser," *Appl. Opt.*, vol. 31, no. 6, pp. 718-731, Feb, 1992.
[17] R. Wei, W. Jiang, and F. K. Tittel., "Single-QCL-based absorption sensor for simultaneous trace-gas detection of $CH_4$ and $N_2O$," *Appl. Phys. B.*, vol. 117, no. 1, pp. 245-251, Apr. 2014.
[18] H. Li, G. B. Rieker, X. Liu, J. B. Jeffries, R. K. Hanson, "Extension of wavelength-modulation spectroscopy to large modulation depth for

diode laser absorption measurements in high-pressure gases," *Appl. Opt.*, vol. 45, no. 5, pp. 1052-1061, Feb. 2006.
[19] G. B. Rieker, J. B. Jeffries, and R. K. Hanson, "Calibration-free wavelength-modulation spectroscopy for measurements of gas temperature and concentration in harsh environments," *Appl. Opt.*, vol. 48, no. 29, pp. 5546-5560, Oct. 2009.
[20] K. Sun, X. Chao, R. Sur, C. S. Goldenstein, J. B. Jeffries, and R. K. Hanson, "Analysis of calibration-free wavelength-scanned wavelength modulation spectroscopy for practical gas sensing using tunable diode lasers," *Meas. Sci. Technol.*, vol. 24, no. 12, pp. 125203-125214, Oct. 2013.
[21] C. S. Goldenstein, C. L. Strand, I. A. Schultz, K. Sun, J. B. Jeffries, and R. K. Hanson, "Fitting of calibration-free scanned-wavelength-modulation spectroscopy spectra for determination of gas properties and absorption lineshapes," *Appl. Opt.*, vol. 53, no. 3, pp. 356-367, Jan. 2014.
[22] C. Yang, L. Mei, H. Deng, Z. Xu, B. Chen, and R. Kan, "Wavelength modulation spectroscopy by employing the first harmonic phase angle method," *Opt. Express.*, vol. 27, no. 9, pp. 12137-12146, Apr. 2019.
[23] C. Zhu, P. Wang, T. Chu, F. Peng, and Y. Sun, "Second Harmonic Phase Angle Method Based on WMS for Background-Free Gas Detection," *IEEE Photonics. J.*, vol. 13, no. 5, pp. 1-6, 2021.
[24] R. Yang, Y. Zhang, "A method of low concentration methane measurement in tunable diode laser absorption spectroscopy and Levenberg-Marquardt algorithm," *Optik.*, vol. 224, pp. 165657-165671, 2020.
[25] J. Li, Z. Peng, Y. Ding, "Wavelength modulation-direct absorption spectroscopy combined with improved experimental strategy for measuring spectroscopic parameters of $H_2O$ transitions near 1.39 µm," *Opt. Laser. Eng.*, vol. 126, pp. 105875-105882, Oct. 2019.
[26] J. Li, H. Deng, J. Sun, B. Yu, H. Fischer, "Simultaneous atmospheric CO, $N_2O$ and $H_2O$ detection using a single quantum cascade laser sensor based on dual-spectroscopy techniques," *Sens. Actuators B Chem.*, vol. 231, pp. 723-732, Mar. 2016.
[27] J. Shao, Y. Huang, L. Dong, Y. Zhang, F.K. Tittel, "Automated rapid blood culture sensor system based on diode laser wavelength-modulation spectroscopy for microbial growth analysis," *Sens. Actuators B Chem.*, vol. 273, pp. 656-663, 2018.
[28] J. Dang, J. Zhang, X. Dong, L. Kong, H. Yu, "A trace $CH_4$ detection system based on DAS calibrated WMS technique," *Spectrochim. Acta A Mol. Biomol. Spectrosc.*, vol. 266, pp. 120418-129424, Sep. 2021.
[29] L. Dong, Y. Yu, C. Li, S. So, F.K. Tittel, "Ppb-level formaldehyde detection using a CW room-temperature interband cascade laser and a miniature dense pattern multipass gas cell," *Opt. Express.*, vol. 23, no. 15, pp. 19821-19830, Jul. 2015.
[30] R. Cui, L. Dong, H. Wu, S. Li, L. Zhang, W. Ma, W. Yin, L. Xiao, S. Jia, and F. K. Tittel, "Highly sensitive and selective CO sensor using a 2.33 µm diode laser and wavelength modulation spectroscopy," *Opt. Express.*, vol. 26, no. 19, pp. 24318-24328, Sep. 2018.

**Tingting Zhang** was born in Dezhou, Shandong, China, in 1997. She is currently pursuing the M.S. degree at the school of physics science and information technology with Liaocheng University. Her current research interests include optical gas sensing devices, optical fiber sensor fabrication and intelligent algorithms.

**Yongjie Sun** was born in Zibao, Shandong, China, in 1997. He is currently pursuing the M.S. degree in physics and technology with University of Jinan. His current research interests include optical fiber sensor fabrication and intelligent algorithms.

**Pengpeng Wang** was born in Jinan, Shandong, China, in 1985. She received the Ph.D. degree from Shandong University in 2014. She is currently with the School of Physics Science and Information Technology, Liaocheng University, Liaocheng, China. Her current research interests include optical fiber sensors and fiber lasers.

**Yufeng Qiu** was born in Zhangjiakou, Hebei, China, in 1996. He is currently pursuing the M.S. degree at the school of physics science and information technology with Liaocheng University. His current research interests include optical fiber sensors and fiber lasers.

**Chenxi Wang** was born in Liaocheng, Shandong, China, in 1999. She is currently pursuing the M.S. degree at the school of physics science and information technology with Liaocheng University. Her current research interests include optical fiber sensor fabrication.

**Xiaohui Du** was born in Weifang, Shandong, China, in 1997. She is currently pursuing the M.S. degree at the school of physics science and information technology with Liaocheng University. Her current research interests include optical fiber sensors and fiber lasers.

**Shaokai Li** was born in Jining, Shandong, China, in 2000. He is currently pursuing the M.S. degree at the school of physics science and information technology with Liaocheng University. His current research interests include optical fiber sensors.

**Haixu Liu** was born in Weifang, Shandong, China, in 1999. He is currently pursuing the M.S. degree at the school of physics science and information technology with Liaocheng University. His current research interests include optical fiber sensor fabrication.

**Tongwei Chu** was born in Liaocheng, Shandong, China, in 1998. He is currently pursuing the Ph.D. degree in electronic engineering with Duke University. His current research interests include optical gas sensing devices, optical sensor fabrication, and engineering applications.

**Cunguang Zhu** was born in Liaocheng, Shandong, China, in 1985. He received the Ph.D. degree in optoelectronic engineering from Shandong University, Jinan, China, in 2015. He is currently with the School of Physics Science and Information Technology, Liaocheng University, Liaocheng. His current research interests include optical gas sensing devices, optical fiber sensor fabrication, and engineering applications.